\documentclass[aps,prd,reprint,superscriptaddress,floatfix]{revtex4-2}

\usepackage[utf8]{inputenc} 
\usepackage[T1]{fontenc}    
\usepackage{hyperref}       
\usepackage{url}            
\usepackage{booktabs}       
\usepackage{amsfonts}       
\usepackage{nicefrac}       
\usepackage{microtype}      
\usepackage{lipsum}		
\usepackage{graphicx}
\usepackage{natbib}
\usepackage{amsmath}
\usepackage{float}
\usepackage{comment}

\usepackage{xcolor}
\newcommand{\carol}[1]{\textcolor{purple}{#1}}
\newcommand{\peter}[1]{\textcolor{orange}{#1}}
\newcommand{\hide}[1]{}
\begin{document}

\title{$w_0$ Is All You Need: Field-Level Matching of General Dark Energy Models \\ to $w_0$CDM Simulations}

\author{Pavel Demidov}
\email{pd2487@nyu.edu}
\affiliation{Center for Cosmology and Particle Physics,
Department of Physics, New York University,
726 Broadway, New York, NY 10003, USA}

\author{Carolina Cuesta-Lazaro}
\affiliation{Center for Cosmology and Particle Physics,
Department of Physics, New York University,
726 Broadway, New York, NY 10003, USA}
\affiliation{Center for Computational Astrophysics, Flatiron Institute, 162 5th Avenue, New York, NY 10010, USA}

\author{Peter L.~Taylor}
\affiliation{Center for Cosmology and AstroParticle Physics (CCAPP), The Ohio State University, Columbus, OH
43210, USA}
\affiliation{Department of Physics, The Ohio State University, Columbus, OH 43210, USA}
\affiliation{Department of Astronomy, The Ohio State University, Columbus, OH 43210, USA}

\begin{abstract}
    We show that clustering in general dynamical dark energy models $w(z)$ can be predicted with constant dark energy $w_0\text{CDM}$ models at the field level via spectral equivalence. The prescription of this method requires equating comoving distances to the last scattering surface between predictor and target models, which is equivalent to requiring equal recent background expansion history averages. Dark energy models with the same distance thus form approximate equivalence classes, with nearly identical matter fields. To test the precision of spectral equivalence beyond the CPL parametrization, we run $N$-body simulations with various dynamical dark energy models, including spline-based $w(z)$, all matched to a single $w_0\text{CDM}$ run. Our measurements show that the power spectra of all matched dark energies agree with the predictor within $0.2\%$ up to $k = 0.5\,h\,{\rm Mpc}^{-1}$ and within $0.6\%$ up to $k = 1\,h\,{\rm Mpc}^{-1}$. We also compare the non-Gaussian information contained in the simulated fields. Bispectra differ by at most $0.26\%$ across all runs up to $k = 0.5\,h\,{\rm Mpc}^{-1}$ and $1.75\%$ up to $k = 1\,h\,{\rm Mpc}^{-1}$, while wavelet scattering transform coefficients are all within $0.05\%$. The decorrelation, $1 - r^2$, between each matched field and the predictor stays below $0.006\%$  up to $k = 1\,h\,{\rm Mpc}^{-1}$.  We run a toy power spectrum analysis in the CPL parametrization demonstrating that a $w_0$-only emulator could produce the same posterior without any knowledge of the full $w_0w_a$ model. This matching prescription thus offers a new tool to constrain dynamical dark energy models with arbitrary equations of state at the field level, without running additional simulation suites.
\end{abstract}

\maketitle

\section{Introduction}

Recent measurements from the Dark Energy Spectroscopic Instrument (DESI) collaboration suggest that dark energy may be evolving~\citep{desi_dr1_bao, desi_dr2_bao}. Explaining this phenomenon naturally poses a serious theoretical challenge. The current approach is purely empirical: the deviations of the time-evolving dark energy equation of state (EOS), $w(z)$, are parametrized via a first-order Taylor expansion in the scale factor, $w(a) = w_0 + w_a (1 - a)$. This is called the Chevallier-Polarski-Linder (CPL) parametrization~\citep{chevallier_2001, Linder_2003}. There is no fundamental reason for dark energy to be described by this particular EOS. Data from Euclid~\citep{Euclid_2025}, the Nancy Grace Roman Space Telescope (Roman)~\citep{Spergel_2015} and the Vera C. Rubin Observatory's Legacy Survey of Space and Time (Rubin-LSST)~\citep{Ivezi_2019} may let us probe broader classes of models or significantly tighten the existing constraints~\citep{Euclid_forecast_2020, Eifler_2021, Xu_2022, lsst_2012}. 

Current techniques are incapable of modeling structure for arbitrary dynamical dark energy models into the nonlinear and non-Gaussian regimes. Usually, this requires running a very expensive $N$-body simulation suite for each new parametrization. Moreover, increasing the number of parameters in the EOS increases the volume of the parameter space exponentially, which will make high-precision emulation of summary statistics infeasible. 

This state of affairs forces the community to look for other solutions. The BACCO project~\citep{Angulo_2021} rescales a small set of simulations to construct nonlinear predictions for other cosmological models, including dynamical dark energies. BACCO achieves $\leq 3\%$ error in the power spectrum all the way up to $k = 5\, h \, \text{Mpc}^{-1}$; however, it was tested only in a narrow range of CPL cosmologies. 

Another approach is to model the evolution of summary statistics as a neural ODE, which can be trained on a broad set of expansion histories, so that a neural network could interpolate between them~\citep{neural_ode}. The idea is elegant, but it is unclear whether existing simulation suites cover sufficiently distinct histories to make this approach applicable to any dynamical dark energy model. It was also tested only for the power spectrum, and it is yet to be shown that this ansatz can work for other summaries.

There is a separate family of methods that explicitly focuses on the background history. One of the first examples demonstrating the importance of recent expansion can be found in~\citep{Linder_2005}. Recent work in this direction is focused on developing this approach. Evolution mapping corrects the history mismatch between two snapshots sharing the same linear power spectrum shape~\citep{Sanchez_2022, Sanchez_2026}. Spectral equivalence~\citep{Francis_2007, Casarini_2009, spec_equivalence_2, spec_equivalence_1} takes another approach: it predicts the power spectrum of $w_0w_a$ models to $\lesssim1\%$ from a $w_0\text{CDM}$ stand-in cosmology with the same distance to last scattering, recently also within the DESI DR2\,+\,CMB posterior~\citep{Chen_2026}.

We argue that spectral equivalence extends beyond the power spectrum and CPL dark energies. We perform $N$-body simulations of cosmologies with exotic shapes of $w(z)$ to stress-test this approach and demonstrate that even in an adversarial case this matching can produce solid predictions for various dark matter clustering statistics, including the bispectrum, the wavelet scattering transform (WST) and the one-point density PDF. We also show that a naive implementation of this method does not generate accurate velocity fields, but it can be corrected via linear theory arguments, which improves our predictions of redshift-space galaxy clustering. To compare the fields even further, we study how halos and voids differ between simulations.

We suggest that this method can be pushed to field-level analyses by demonstrating that the Fourier modes of each matched simulation and of the predictor agree in amplitude and phase to better than $1\%$ for $k \leq 1\,h\,{\rm Mpc}^{-1}$, with a cross-correlation coefficient $r > 0.9999$.

The paper is structured as follows. In Sec.~\ref{sec:headings} we describe the logic behind paying close attention to recent expansion history and how evolution mapping uses it to make predictions. We explain why it cannot scale to field-level analyses and why spectral equivalence is a good alternative despite requiring more simulations. In Sec.~\ref{sec:N_body} we test this framework with $N$-body simulations at $z = 0$ and demonstrate that the mapping holds for the matter density field itself and, at $k\leq 1\,h\,{\rm Mpc}^{-1}$, predicts statistics beyond the power spectrum, such as the one-point PDF outside of its tails and the WST, within $1\%$, and the bispectrum to within $1.75\%$. The agreement also extends to velocities after a linear rescaling, to halo and void abundances, and to galaxy clustering. In Sec.~\ref{sec:inference} we perform a toy analysis where we use spectral equivalence to sample a CPL posterior with an emulator and compare it against a direct sampling with a full $w_0w_a$ model.

\section{Methodology}
\label{sec:headings}

We aim to find approximate equivalence classes of dynamical dark energy models, $w(z)$, whose nonlinear matter and galaxy clustering at a given redshift can be reproduced, at the field level, by a constant dark energy $w_0\text{CDM}$ cosmology through an adjustment of $w_0$ and of the amplitude of primordial fluctuations.

We first demonstrate that, at a fixed moment of time, $\mu(a) = \Omega_m(a) / f^2(a)$ is the only background history parameter of importance in the $N$-body equations of motion, if linear growth is used as a clock. This suggests that matter and its summaries can depend on the background through $\mu$ and its history. We connect this statement with previous work on evolution mapping, which shows that only averaged recent history matters for nonlinear predictions. We discuss why we may want a different solution, especially if field-level information is required, and present the spectral equivalence method as an alternative at the mild cost of requiring $w_0\text{CDM}$ emulators to be available. We also show that these two methods are not as different as they seem.

\hide{Our goal is to construct an approximate predictor of dark matter properties for some dynamic dark energy model $w(z)$ using existing $w_0\text{CDM}$ simulation suites. This means that we can assume there is a simulator or an emulator of some summary statistic of interest $\mathbf{s}(\text{DM})$ (including an overdensity/velocity field) that can be called for any non-dynamic dark energy cosmological parameters. In this work, we explore the path of substituting some cosmological model $M$ with a constant dark energy proxy $M_\text{pred}$, generalizing previous results (citations) in a single framework and extending it to a broader class of models.}

\subsection{Recent Expansion History Average}

Here, we assume that dark energy contributes only to the background expansion, i.e. that it is smooth on the scales we simulate and does not cluster. We focus on how dark energy affects both the linear growth history and the background expansion. Using the growth factor as the time variable, $\eta = \ln D$, the $N$-body equations of motion read 

\[
\begin{cases}
  \mathbf{x}'' + \left( \frac{3\Omega_m}{2f^2} -1 \right)\mathbf{x}' = - \frac{3\Omega_m}{2f^2} \nabla \psi, \\
  \nabla^2 \psi = \delta.
  \label{eq:Nbody}
\end{cases}
\]

Here $\mathbf{x}$ is the comoving position of a particle, primes denote derivatives with respect to $\eta \equiv \ln D$, with $D$ the linear growth factor, $\Omega_m(\eta)$ is the matter density parameter, $f \equiv d\ln D/d\ln a$ is the linear growth rate, $\delta \equiv \rho_m/\bar{\rho}_m - 1$ is the matter density contrast, and $\psi \equiv \phi/(4\pi G\bar{\rho}_m a^2)$ is the peculiar gravitational potential rescaled so that the Poisson equation carries no dependence on cosmology.

The most important observation is that in this parametrization matter experiences all background effects through a single term $\mu = \Omega_m/f^2$. Two cosmologies with matching $\mu$ would therefore lead to the same particle trajectories. The evolution mapping framework leverages this fact~\citep{Sanchez_2022, Sanchez_2026}. For two cosmologies, possibly at different redshifts, that share the same RMS of linear density fluctuations smoothed over spheres of radius $12 \, \text{Mpc}$, $\sigma_{12}$ (rather than the standard $8\, h^{-1} \, \text{Mpc}$), and other shape parameters ($n_s$, $\omega_b$, $\omega_m$), the mismatch of background expansion histories is quantified with

\begin{equation}
    \tilde{x}(\eta) = \int\limits_{-\infty}^\eta G(\eta') \mu(\eta') {\rm d}\eta',
    \label{eq:x_of_tau}
\end{equation}
where $G(\eta)$ is a kernel selected differently to map different summary statistics. This is used to estimate a chosen summary, $S$, via a linear correction

\begin{equation}
    S \approx S_\text{pred} \left(1 + \frac{\partial R}{\partial \tilde{x}} \Delta\tilde{x} \right),
    \label{eq:EM_corr}
\end{equation}
where $S_\text{pred}$ is a summary from a cosmology that shares all shape parameters, $\partial R/\partial \tilde{x}$ is a correction term calibrated to simulations and $\Delta\tilde{x}$ is the difference in recent history averages due to background differences.

In general, there is no first-principles prescription that fixes the shape of the kernel $G(\eta)$. It is usually parametrized by a half-Gaussian centered on the time of interest, $\eta$, with some width $\sigma$ set empirically. Given that different summary statistics require different predictors and different correction terms, this approach does not extend to the field level. Instead, we find that the spectral equivalence method, while being similar in spirit, produces accurate predictions of various summary statistics for a wide class of dynamical dark energies $w(z)$ by matching to $w_0\text{CDM}$ simulations, without having to train any corrections.

\subsection{The Spectral Equivalence Method and Approximate Equivalence Classes}
\label{sec:spectral_equivalence}
\hide{Usually, spectral equivalence method follows a similar formulation to ours, but instead of averaging $\mu(\eta)$ it requests equality of comoving distances from the redshift of interest $z$ to the surface of last scattering. Since we study only late time dark energy models, we can safely set the upper limit to infinity:}

In our setup, we look to match dynamic dark energy targets $w(z)$ with any cosmology from the $w_0\text{CDM}$ family. First of all, we equate the linear scales of the predictor and the target at the same redshift by adjusting the predictor's primordial amplitude $A_{s,\text{pred}}$ using linear perturbation theory:

\hide{\begin{equation}
    {\sigma_{8}}_{\text{pred}}\frac{D_{\text{M}_\text{pred}}(z)}{D_{\text{M}_\text{pred}}(0)} = \sigma_{8}\frac{D_{\text{M}}(z)}{D_{\text{M}}(0)}.
    \label{eq:linear_equality_old}
\end{equation}
}

\begin{equation}
    A_{s,\text{pred}} = A_s \frac{D^2(z)}{D_\text{pred}^2(z)}.
    \label{eq:linear_equality}
\end{equation}
Here, $D(z)$ is the linear growth factor of the target $w(z)$ cosmology, $A_s$ is its primordial amplitude and the subscript pred denotes the same quantities of the $w_0\text{CDM}$ predictor. Both growth factors are normalized to $D = a$ deep in matter domination.

\hide{\carol{Here, $D_{\rm M}(z)$ is the linear
growth factor of cosmology ${\rm M}$, so that each side of
Eq.~\eqref{eq:linear_equality} is the linear amplitude at the redshift of
interest, $\sigma_8(z)$, in the predictor (left) and in the target cosmology
(right). In practice we impose this condition by adjusting the primordial amplitude $A_s$ of the predictor.}}

In general, the background history of an arbitrary $w_0\text{CDM}$ cosmology does not need to have anything in common with that of another $w(z)$ model. The spectral equivalence approach solves this by adjusting $w_0^\text{pred}$ of the predictor to match the comoving distance to the surface of last scattering $d_\text{LSS}(z)$ of the target

\begin{equation}
    \int\limits_z^{z_\star} \frac{{\rm d}z'}{H(z')} = \int\limits_z^{z_\star} \frac{{\rm d}z'}{H_\text{pred}(z')},
    \label{eq:dLSS_equality}
\end{equation}
where $H(z)$ is the Hubble parameter of the target model $w(z)$ and the subscript pred, again, denotes the predictor's quantity. The upper limit of integration is usually set to the redshift of the last scattering surface $z_\star \approx 1090$, but there is nothing specific about it; we follow the conventions from previous work on this method.

The method proceeds as follows. First, find the predictor dark energy model by solving~\eqref{eq:dLSS_equality} for $w_0^\text{pred}$. Then adjust $A_{s,\text{pred}}$ such that~\eqref{eq:linear_equality} holds. Keep all other cosmological parameters intact. Two equations are solved one after another, rather than jointly. 

At first glance, the condition in Eq.~\eqref{eq:dLSS_equality} does not seem to be related to what evolution mapping does. However, the comoving distance integral can be written as

\begin{equation}
    \int\limits_z^{z_\star} \frac{{\rm d}z'}{H(z')} = \int\limits_{-\infty}^\eta \frac{f(\eta')}{a(\eta')H(\eta')\Omega_m(\eta')} \mu(\eta') {\rm d}\eta',
\end{equation}
which turns the condition in Eq.~\eqref{eq:dLSS_equality} into a requirement of the same average background histories, quantified as in Eq.~\eqref{eq:x_of_tau}, with a single kernel $G(\eta) = \frac{f(\eta)}{a(\eta) H(\eta) \Omega_m(\eta)}$ for $\eta > \eta_\star$ and zero otherwise. Since $\frac{d\ln G(\eta)}{d\eta} = \frac{1}{2} + \frac{3}{2}\left(\mu - 1 \right) > 0$ for all reasonable dark energy models, the kernel of the spectral equivalence method attends mostly to late times.

It has been reported that the spectral equivalence method provides power spectrum predictions for CPL cosmologies at the 1\% level at scales of $k \leq 1 \, h \,\text{Mpc}^{-1}$, see~\citep{Chen_2026}. Here, we argue that this background matching approach can be used beyond CPL cosmologies, so it can make predictions for any $w(z)$. Moreover, its predictive power is not limited to the power spectrum. 

In Sec.~\ref{sec:N_body} we show that it extends to various dark matter summaries and galaxy clustering. Since it only requires a single redshift of a single simulation to make all predictions for another snapshot, spectral equivalence may also be used to analyze fields. This extends the analysis possibilities of existing data without requiring any new simulation suites to train emulators.

This matching approach can be thought of as slicing all dark energy equations of state $w(z)$ into equivalence classes parametrized with a single $w_0^\text{pred}$, although the equivalence is approximate. Compared to evolution mapping, this approach needs more simulations to make predictions: any $w_0\text{CDM}$ cosmology should be available for the matching. But the community has already paid the simulation cost, and there are several publicly available simulation suites providing good coverage of these cosmologies, see CosmoGridV1~\cite{cosmogrid} or Gower Street~\citep{gower_street}. 

\begin{table*}[t]
  \begin{ruledtabular}
    \begin{tabular}{lcccccc}
      Simulation name & $10^9A_s$ & $\sigma_8$ & $w_0$ & $w_a$ & Fluid + PPF ICs  & $d_\text{LSS} / d_{\text{LSS},\text{pred}}$ \\
      \hline
      DESI mean & $2.19020$ & 0.828 & -0.42 & -1.75 & No & 1 \\
      DESI mean (fluid + PPF ICs) & $2.19020$ & 0.828 & -0.42 & -1.75 &  Yes & 1 \\
      $w_0\text{CDM}$ predictor & $2.24930$ & 0.828 & -0.8818 & 0 & No & 1\\
      $w(z)$ & $2.20036$ & 0.828 & --- & --- & No & 1 \\
      adversarial $w_\text{adv}(z)$ & $2.29932$ & 0.828 & --- & --- & No & 1 \\
      Naive guess & $4.20796$ & 0.828 & -0.42 & 0 & Yes & 0.917 \\
    \end{tabular}
  \end{ruledtabular}
  \caption{\textbf{Simulations used in this work.} All simulations share the same initial random seed and differ only in the quantities listed. $A_s$ is set in each run so that all have the same linear amplitude at $z=0$, $\sigma_8 = 0.828$, as required by Eq.~\eqref{eq:linear_equality}. The last column gives the comoving distance from $z=0$ to last scattering relative to that of the predictor, which is unity for the runs matched through Eq.~\eqref{eq:dLSS_equality}. ``Fluid + PPF ICs'' indicates whether dark energy perturbations are included in the initial conditions.}
  \label{table:sims}

\end{table*}

\begin{figure*}
    \centering
    \includegraphics[width=1.0\linewidth]{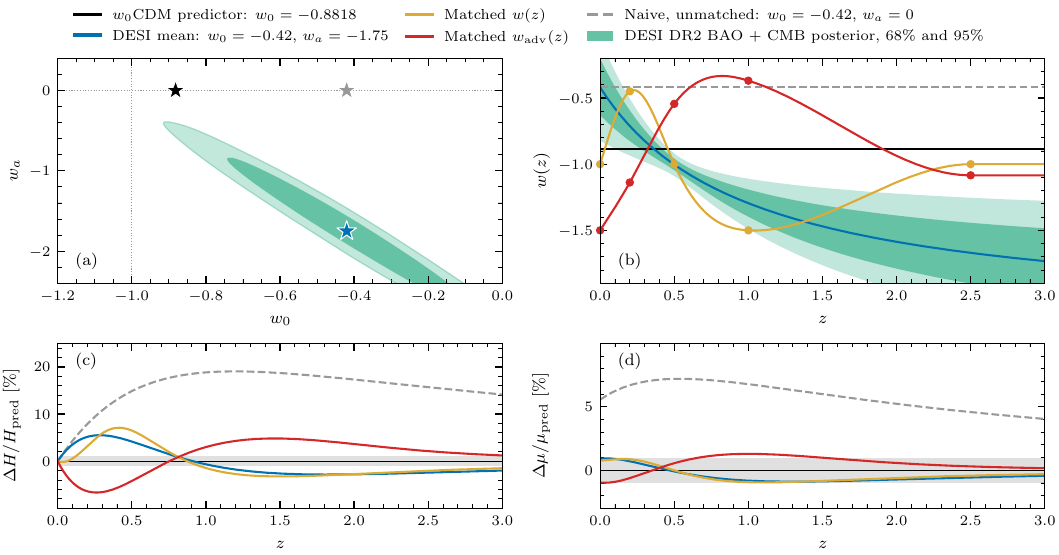}
    \caption{\textbf{Widely different dark energy histories share one $w_0$CDM predictor cosmology.} All simulations in this work have the same comoving distance from $z=0$ to last scattering, $d_{\rm LSS}(0)$, Eq.~\eqref{eq:dLSS_equality}, except the naive guess ($w_0=-0.42$, $w_a=0$). (a) DESI\,+\,CMB posterior in the CPL plane (68\% and 95\% contours) and its mean (blue). \hide{Gray lines connect CPL models with the same $d_{\rm LSS}(z)$ as the DESI mean, for several redshifts, highlighting the dependence of the predictor cosmology on redshift.} We match at $z=0$ \hide{(thick line)}, giving the predictor $w_0 = -0.8818$ used throughout (black). The naive guess (gray) keeps $w_0$ and sets $w_a=0$. (b) Equation of state of each simulated model, with the DESI\,+\,CMB posterior band overlaid. The two spline models (dots mark the five nodes) test the matching outside CPL; the adversarial one maximizes the deviation in $H(z)$ from the DESI mean at fixed $d_{\rm LSS}(0)$. (c, d) Relative difference in $H(z)$ and $\mu(z)=\Omega_m/f^2$ between each simulation and the predictor; the gray band marks $\pm1\%$. Despite their different histories, the matched models stay within $\sim$7\% in $H$ and $\sim$1\% in $\mu$, against $\sim$19\% and $\sim$7\% for the naive guess.}
    \label{fig:DE_models}
\end{figure*}

\section{\texorpdfstring{$N$}{N}-body results}
\label{sec:N_body}

\subsection{Description of Simulations}
We test the spectral equivalence matching of Sec.~\ref{sec:spectral_equivalence} with a suite of $N$-body simulations. This lets us probe nonlinear scales and a broad range of summary statistics. The target cosmologies are: (i) one CPL model, placed at the mean of the DESI + CMB posterior, and (ii,iii) two spline-based $w(z)$ models included to show that the method works well beyond CPL parametrizations of dark energy. All three simulations share the same comoving distance from $z=0$ to last scattering, given by Eq.~\eqref{eq:dLSS_equality}. Hence, all can be compared to the same $w_0 = -0.8818$ predictor cosmology at $z=0$. We do not test any other redshifts, since this would require running different matching simulations.

The two spline models (see Fig.~\ref{fig:DE_models}) are defined by cubic interpolation of $w(z)$ through five nodes, with $w$ held constant beyond the last node. The first, denoted $w(z)$, follows $\Lambda\text{CDM}$ above $z = 2.5$ and at lower redshift oscillates outside the $2\sigma$ region of the DESI + CMB posterior. The second, which we denote adversarial $w_\text{adv}(z)$, was constructed in our attempt to break the method. It was optimized to maximize the deviation in $H(z)$ from the DESI mean model, at the fixed comoving distance defined in Eq.~\eqref{eq:dLSS_equality}, subject to $w(z) \leq -1/3$ at all $z$, knots no lower than $-2$ (and $-1.5$ at $z = 0$), and $\rho_\text{DE}(z = 127) \leq \rho_\text{DE}(0)$, so that dark energy always accelerates the expansion and is negligible at early times. Maximizing the deviation in $\mu(\eta)$ instead of $H(z)$ would lead to a similar adversarial equation of state. This model should be close to the worst case for the matching method. By construction, both $w(z)$ models have the same comoving distance as the DESI mean model and should therefore correspond to the same predictor cosmology, $w_0 = -0.8818$.

To demonstrate the effectiveness of the method, we also run a naive-guess simulation with $w_0 = -0.42$, $w_a = 0$. Its amplitude is still adjusted to satisfy Eq.~\eqref{eq:linear_equality}, but its expansion history is not matched through Eq.~\eqref{eq:dLSS_equality}. This simulation helps us isolate how much of the agreement comes from matching linear scales alone, versus matching the expansion history.

Fig.~\ref{fig:DE_models} shows the simulated dark energy models and their equations of state, overlaid on the DESI posterior obtained with the CPL parametrization. Panels (c) and (d) highlight the relative differences in $H(z)$ and $\mu(z)$, showing that the matched models stay within $\sim$7\% in $H$ and $\sim$1\% in $\mu$, against $\sim$19\% and $\sim$7\% for the naive guess.

All simulations were run with the TreePM code \textsc{Gadget-III} \citep{gadget}, in the same configuration as the high-resolution Quijote simulations \citep{Quijote}. Each follows $1024^3$ dark matter particles in a periodic box of side $1\,h^{-1}\,{\rm Gpc}$, corresponding to a particle mass of $8.2\times10^{10}\,h^{-1}\,M_\odot$, with a gravitational softening length of $1/40$ of the mean interparticle separation, $\simeq 25\,h^{-1}\,{\rm kpc}$. Initial conditions were generated at $z=127$
with \textsc{monofonIC} \citep{monofonIC} using 2LPT, and all runs share the same random seed to match cosmic variance. Dark energy enters the $N$-body evolution only through the expansion rate, $H(z)$, tabulated for each $w(z)$. 

All runs share the fiducial Quijote parameters, $\Omega_m = 0.3175$, $\Omega_b = 0.049$, $h = 0.6711$, $n_s = 0.9624$ and massless neutrinos, and differ only in the dark energy equation of state and in $A_s$, which is set so that $\sigma_8 = 0.828$ at $z=0$ in every run (Table~\ref{table:sims}).

Our first runs included dark energy perturbations in the initial conditions, through the fluid and parametrized post-Friedmann (PPF) description. For a sound speed $c_s = 1$, these perturbations are suppressed inside the horizon and should only appear at scales $k \ll aH/c_s \sim 2.2\times10^{-3}\,h\,{\rm Mpc}^{-1}$ at $z = 127$, below even the fundamental mode of our boxes. Their effect is therefore confined to the largest scales and expected to be very small. We found, however, that there is a non-zero mismatch in dark matter clustering on the largest modes of the box, 0.2\% at $k < 10^{-2} \,  h \, \text{Mpc}^{-1}$ in the ICs, which is visible in halo clustering at $z = 0$ but negligible in current data analysis. Since we aim to test the matching method at the level of the background only, all other runs were done from the same $\Lambda$CDM initial conditions, without dark energy perturbations. We keep the original DESI mean run with and without perturbations to quantify their impact in Appendix~\ref{ap:de_perturb}. The naive guess was the only other run with perturbations included.

\subsection{Dark Matter Clustering}

Our simulation suite lets us probe a wide range of dark matter clustering summary statistics, beyond the power spectrum. Given the resolution of each run, we limit our study to scales of $k \leq 1 \, h \, \text{Mpc}^{-1}$. To start, we compare some commonly used clustering statistics: the one-point PDF, the power spectrum, the bispectrum and the WST.

All correlation functions were estimated on a $1024^3$ grid with a cloud-in-cell (CIC) mass assignment scheme. We use Pylians~\citep{Pylians} to measure the power spectrum and PolyBin3D~\citep{polybin3d} for the bispectrum. For the WST measurements, we closely follow the choices used for large-scale structure data analyses performed in~\cite{Valogiannis2022, Valogiannis2024}: we set $J = 4$, $L = 4$, $q = 0.8$ and $\sigma_0 = 0.8$ on grids with triangular-shaped-cloud (TSC) mass assignment. To increase sensitivity to smaller scales, we use higher-resolution $256^3$ grids with a cell size of $3.9 \, h^{-1} \, \text{Mpc}$.

\begin{figure*}
    \centering
    \includegraphics[width=\textwidth]{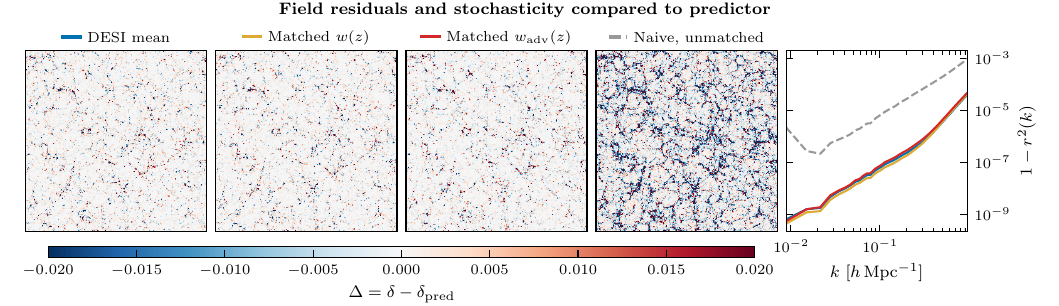}
   \caption{\textbf{Field-level comparison with the $w_0$CDM predictor at $z=0$.} The maps show the residual $\Delta = \delta - \delta_{\rm pred}$ of the projected density in a $20\,h^{-1}\,{\rm Mpc}$ slab across the full box, on $3.9\,h^{-1}\,{\rm Mpc}$ cells. The matched models differ from the predictor only around the densest structures, while the residual of the naive guess traces the whole cosmic web and saturates the color scale. The right panel shows the decorrelation $1 - r^2(k)$, the fraction of the power that no transfer function of $\delta_{\rm pred}$ can reproduce. It stays below $6\times10^{-5}$ for the matched models up to $k = 1\,h\,{\rm Mpc}^{-1}$, against $\sim10^{-3}$ for the naive guess.} 
    \label{fig:field_residuals}
\end{figure*}

\begin{figure*}[p]
    \centering
    \includegraphics[width=\textwidth]{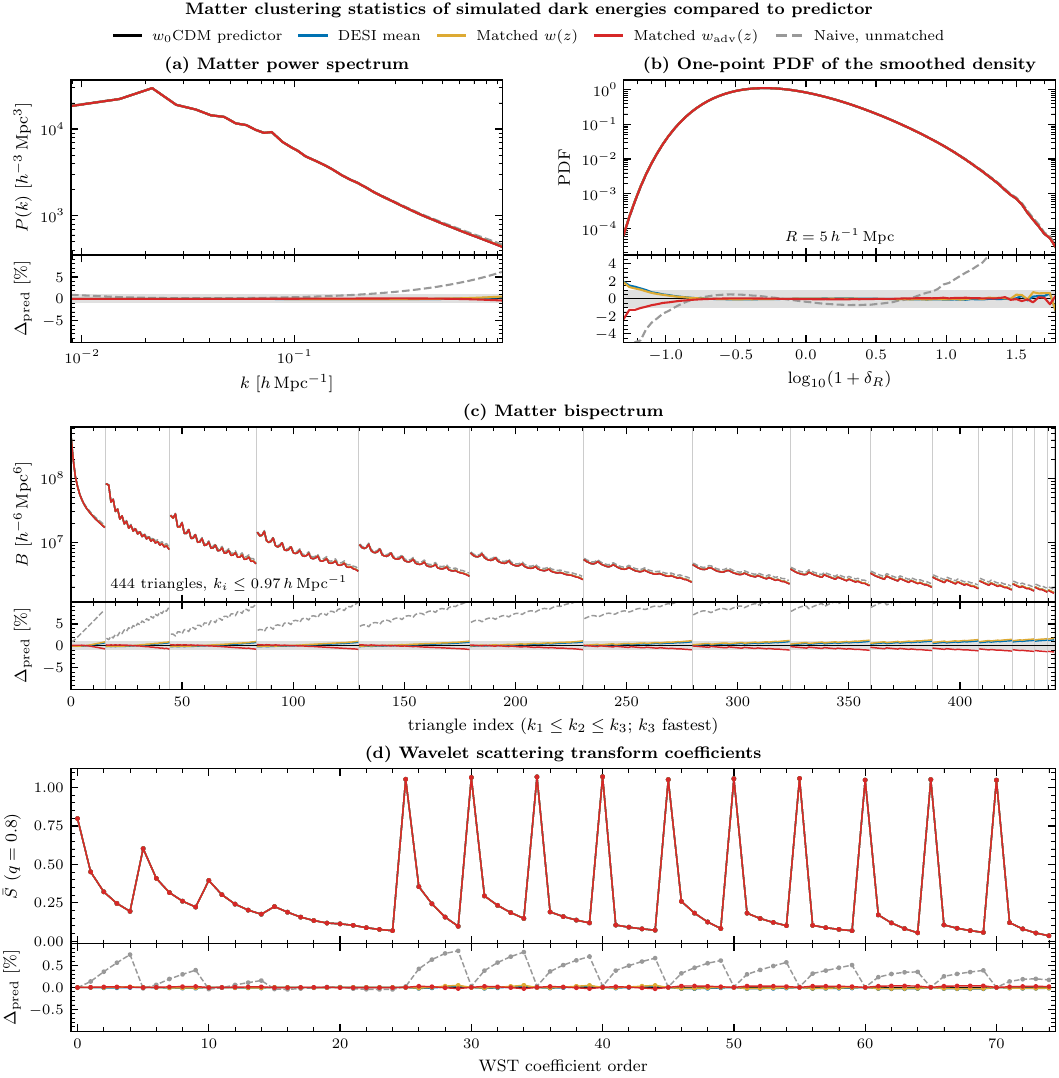}
    \caption{\textbf{Dark matter clustering summary statistics comparison of the matched cosmologies.} (a) Matter power spectra of all matched dark energies are within $0.2\%$ up to $k = 0.5 \, h \, \text{Mpc}^{-1}$ and within $0.6\%$ up to $k = 1 \, h \, \text{Mpc}^{-1}$ compared to the predictor $w_0\text{CDM}$ cosmology. (b) One-point PDFs of the densities smoothed with a top-hat filter with $R = 5 \, h^{-1} \,\text{Mpc}$ agree with the predictor to within $0.2\%$ for $-0.8 < \log_{10}(1 + \delta_R) < 1.5$, a range containing $97.3\%$ of cells. The differences grow to $2.3\%$ only in the emptiest, and therefore noisiest, cells. The naive model (dashed gray) has too few of the emptiest cells and too many of the densest: its PDF deviates by $9\%$ and $15\%$ at the edges of the plotted range. (c) Bispectrum mismatches reach $1.75\%$ for the worst-case equilateral triangle $k_1 = k_2 = k_3 = 0.97 \, h \, \text{Mpc}^{-1}$ in comparison to the predictor, while the naive guess misses the predictor by $14\%$. Staying under $k \leq 0.5 \, h \, \text{Mpc}^{-1}$ reduces errors of all three matched models down to $0.26\%$, while the naive guess is off by $7.1\%$. These three summary statistics were computed on a $1024^3$ grid with CIC mass assignment. (d) WST coefficients measured on a $256^3$ grid with TSC mass assignment. The wavelet configuration is set to $\sigma_0 = 0.8, \: J = 4,\: L = 4$ with the power $q = 0.8$ to keep it similar to standard pipelines. The three matched simulations are within $0.05\%$ of the predictor, while the naive guess deviates by $0.84\%$ in the worst bin. When compared to cosmic noise from the fiducial Quijote simulations, these numbers translate into error-to-noise ratios of up to $0.25$ and $10$, respectively.}
    \label{fig:clustering}
\end{figure*}

The field-level results are presented in Fig.~\ref{fig:field_residuals} while the clustering statistics results are shown in Fig.~\ref{fig:clustering}. While the naive guess simulation deviates significantly from all other runs (since its dark energy was not matched with spectral equivalence), every other cosmology stays very close to our $w_0\text{CDM}$ predictor. Among the matched simulations, the highest error in the power spectrum is obtained by the $w(z)$ simulation: its error grows to $0.6\%$ at $k = 1.0 \, h \, \text{Mpc}^{-1}$, while the naive guess simulation is off by $6.2\%$ at the same scales. 

The same pattern repeats for all other summaries: the worst matched cosmology is off by $1.75\%$ in the bispectrum at the triangle with $k_1 = k_2 = k_3 = 0.97 \, h \, \text{Mpc}^{-1}$ compared to a $14\%$ error in the naive guess. 

The one-point PDF also demonstrates that, even in the inherently noisy tails, errors stay within $2.3\%$ for a smoothing scale of $R = 5 \, h^{-1} \, \text{Mpc}$, while the naive guess cosmology has errors over $15\%$.

Differences in WST coefficients come out much smaller. Their size alone is not informative, however, because the sample variance of these coefficients is also very small, so even sub-percent deviations can be statistically significant. To put the WST errors into perspective, we estimate the cosmic variance of this summary with 500 realizations of the fiducial Quijote simulations with $512^3$ particles. The ratio of errors to noise stays below $0.25\sigma$ for all matched simulations, while for the naive guess it reaches $10\sigma$.

Overall, we find that the adversarial equation of state is matched as accurately as the CPL cosmology, despite being built to break the method. This suggests that the errors of the matching are bounded over the space of smooth dark energy histories, and could therefore be quantified with a small set of simulations. We leave this for future work.

Finally, we test the mapping at the field level. Because all simulations share the same initial phases, their density fields can be compared mode by mode, through the residual $\Delta(\mathbf{x}) \equiv \delta(\mathbf{x}) - \delta_{\rm pred}(\mathbf{x})$ between each target simulation and the predictor. We quantify it with their decorrelation $1 - r^2(k)$, where $r$ is the cross-correlation coefficient of the two fields. It stays below $0.006\%$ for all matched simulations and reaches $0.1\%$ for the naive guess run. Both field residuals and $1-r^2$ are presented on Fig.~\ref{fig:field_residuals}. The residuals of the matched models are confined to the surroundings of collapsed structures. We also check the error power spectra, which stay below $\Delta P / P \sim (5-6) \times 10^{-5}$. The naive guess error power spectrum is $~\sim40$ times larger tat $k = 1\,h\,{\rm Mpc}^{-1}$.

\subsection{Dark Matter Velocities} \label{sec:velocities}

The velocity field of dark matter is another important channel of information. Not only can it be measured directly in peculiar velocity surveys, but it also affects clustering measurements through redshift-space distortions, the kinematic Sunyaev--Zel'dovich (kSZ) effect on Cosmic Microwave Background photons as they travel towards us, and other cosmological probes. A naive comparison of velocities in the matched simulations would clearly fail, even on large scales, as the following argument shows. In linear theory, the velocity field can be expressed

\begin{equation}
    \mathbf{v}(\mathbf{k}) = i\,aHf\,\frac{\mathbf{k}}{k^{2}}\,\delta(\mathbf{k})
\end{equation}
where the prefactor $aHf$, in general, would not be reproduced by our matching cosmology, which would lead to different velocity fields. We propose a correction that can be applied to the predictor's snapshots based on this consideration. First, we correct each halo's bulk velocity, the mean velocity of all particles assigned to the halo, with

\begin{equation}
    \mathbf{v}_\text{bulk} \to s \mathbf{v}_\text{bulk},
    \label{ref:bulk_correction}
\end{equation}
where $s = fH / f_\text{pred}H_\text{pred}$. The scale factor does not appear in the correction, since we use the same moment of time to perform the matching. The core velocity, the average velocity of the dark matter particles within the innermost 10\% of the halo radius, is then shifted such that the relative motion of the core with respect to the bulk motion stays the same:

\begin{equation}
    \mathbf{v}_\text{core} \to \mathbf{v}_\text{core} + \left( s - 1 \right) \mathbf{v}_\text{bulk}.
    \label{ref:core_correction}
\end{equation}
Since our correction accounts for the large-scale mismatch, we do not apply this modification to particles decoupled from the cosmic expansion. \hide{Virialized velocities are set by the local potential, which should be the same in both cosmologies.} Virialized velocities strongly depend on the local potential rather than on the cosmic expansion, so a background-only correction is irrelevant for them. Those particles that do not have a host get their velocities adjusted as in Eq.~\eqref{ref:bulk_correction}.

\begin{figure}[!htbp]
    \centering
    \includegraphics[width=1.0\linewidth]{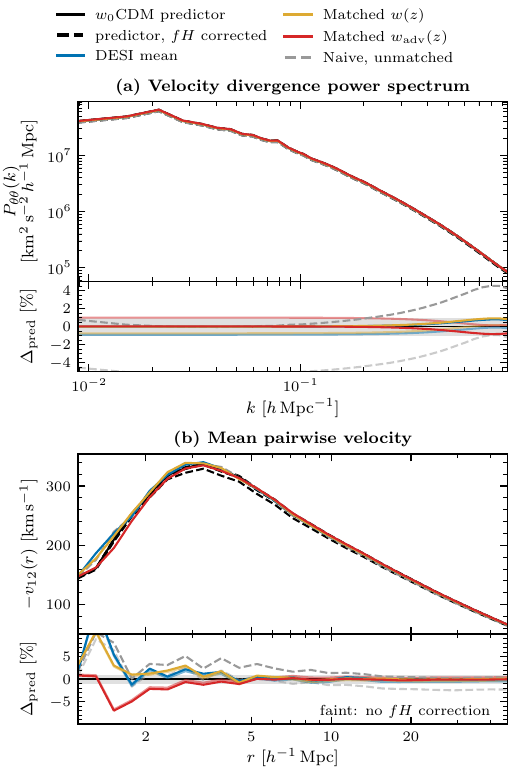}
    \caption{\textbf{Large-scale mismatch in the velocity fields can be corrected via linear theory.} (a) The velocity divergence power spectrum of matter shows a 1\% mismatch on large scales (faint lines), which can be corrected with the $fH$-rescaling of particle velocities (solid lines). The corrected spectra deviate by less than 1\% from the predictor at the smallest scales we consider, $k \sim 1\,h\,{\rm Mpc}^{-1}$. The naive guess (gray) also benefits from the correction, but nonlinear effects start growing and the mismatch reaches 4.5\% at $k \approx 0.9\,h\,{\rm Mpc}^{-1}$. The largest-scale bin shows an upward deviation, mimicking a similar deviation of the matter power spectrum for this simulation, caused by the PPF formalism used by the IC generator for this simulation. (b) The mean pairwise velocity of dark matter particles also shows that large-scale mismatches are removed by the correction, especially for the naive guess simulation (faint vs solid gray dashed lines). All matched simulations stay within 1\% down to $r = 5\,h^{-1}\,{\rm Mpc}$, while the corrected naive guess (gray) reaches the same error threshold at $r = 14\,h^{-1}\,{\rm Mpc}$.}
    \label{fig:vel}
\end{figure}

This correction is different for different cosmologies that we want to match with the predictor, and $s$ changes depending on the dark energy model. Both corrected and uncorrected velocity divergence power spectra and mean pairwise velocities (measured on a subsample of $10^6$ particles) are presented in Fig.~\ref{fig:vel}. Without the correction, the offsets reach $\sim1\%$ in the velocity divergence spectrum $P_{\theta\theta}$ and $\sim0.5\%$ in the mean pairwise velocity. With it, the matched models agree with the predictor to $0.15\%$ at $k \leq 0.3\,h\,{\rm Mpc}^{-1}$ in the spectrum and to $0.41\%$ in the mean pairwise velocity for $r \geq 10\,h^{-1}\,{\rm Mpc}$. At smaller scales, the correction stops helping due to nonlinear evolution effects that are not accounted for by our linear argument.

\subsection{Halos and Voids}
Throughout cosmic evolution, dark matter collapses into halos, leaving behind underdense regions that become voids. Both of these objects are more directly linked to observations than dark matter in general: galaxies are assumed to trace halos, and therefore we study how halos and voids differ between our matched simulations. 

\begin{figure}
    \centering
    \includegraphics[width=1.0\linewidth]{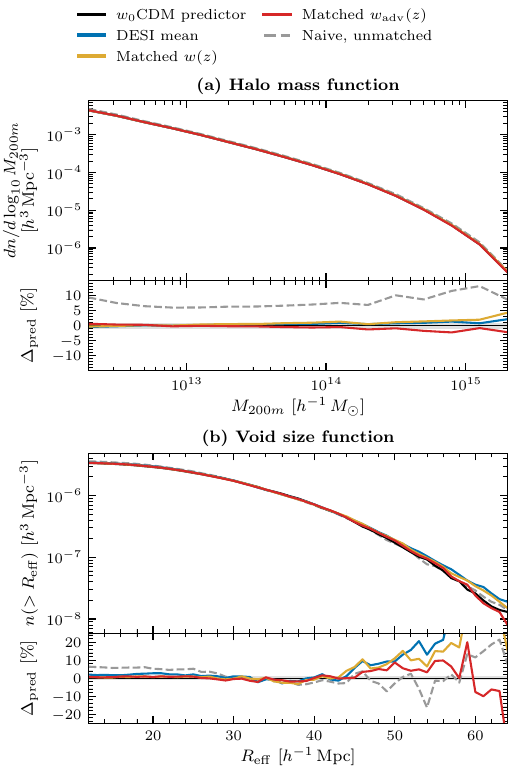}
    \caption{\textbf{Halo mass function and void size function are matched by the predictor.} (a) Halo mass functions of all simulations with relative errors compared to the predictor. Errors of the matched cosmologies stay below $2\%$ up to $M = 10^{14}\,h^{-1}\,M_\odot$ and below $4.5\%$ for the whole range reaching $M = 10^{15.3}\,h^{-1}\,M_\odot$. The naive guess (not matched) misses the target by $6$--$9\%$ up to $M = 10^{14.3}\,h^{-1}\,M_\odot$ and by up to $13.2\%$ for heavier halos. (b) Void size functions (VSF) of all simulations, again, with relative errors compared to the predictor. Relative errors of the matched cosmologies stay within $3.2\%$ for $R_\text{eff} \leq 40\,h^{-1}\,{\rm Mpc}$, while larger voids appear to mismatch strongly due to the very low sample size of only $\sim700$ voids. The naive guess struggles to predict the correct VSF even at the well-sampled effective radii, where the matched simulations perform best.}
    \label{fig:hmf_vsf}
\end{figure}

First, we do a population-wide comparison through the halo mass function and the void size function, shown in Fig.~\ref{fig:hmf_vsf}. In each simulation we identify halos with Rockstar~\citep{rockstar}, with an overdensity threshold of $\Delta = 200$. For the halo mass function, we use all halos that have at least 20 particles in them. Voids are found with VIDE~\citep{sutter2015vide} using halos with at least 100 particles as tracers. 

Then, we study discrepancies on the halo-by-halo level, by constructing a one-to-one map between halos in each simulation and halos in our reference $w_0\text{CDM}$ predictor. Since all simulations are initialized with the same seed, particle IDs can be used to index the same matter particles across all simulations. We define two halos to be matched if they share more than 50\% of each other's particles. This procedure is symmetric under permutation of the catalogs. To avoid collisions between halos, if a particle belongs to multiple halos at the same time, it gets assigned to the heaviest available host. This rule prevents the matching procedure from identifying multiple partners for a single halo. We focus on well-resolved halos with $N \geq 100$ particles; in this case, 99.7\% of objects end up being matched to some halo from the $w_0\text{CDM}$ predictor in all simulations except for the naive guess $w_0 = -0.42$, which ends up having a lower success rate of 99.5\%.

\begin{figure*}
    \centering
    \includegraphics[width=\textwidth]{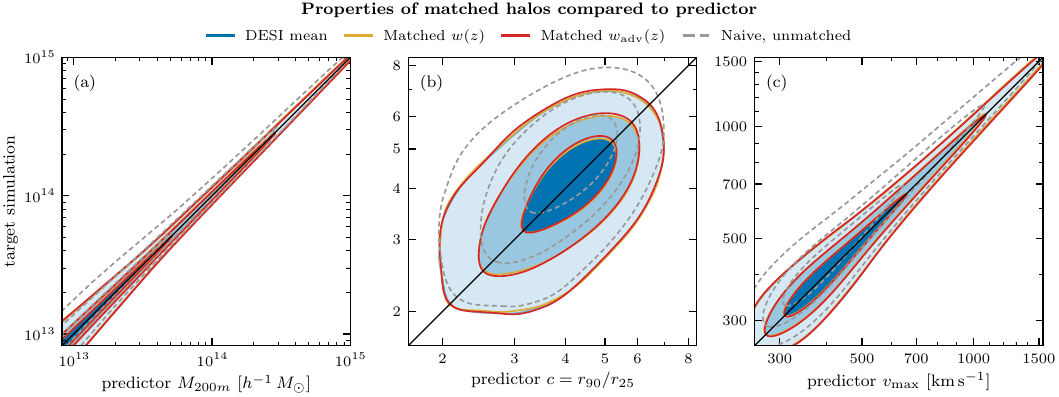}
     \caption{\textbf{Matched halos agree with their predictor counterparts on average.} Mass $M_{200m}$, concentration $c = r_{90}/r_{25}$ and $v_\text{max}$ of halos with $N \geq 100$ particles in each simulation (vertical axis) against their matched counterparts in the predictor (horizontal axis). Contours enclose 68, 95 and 99.7\% of the pairs. For the matched cosmologies, the median offset of each property is at most $0.003$\,dex and the scatter (standard deviation) is at most $0.022$, $0.065$ and $0.023$\,dex in (a), (b) and (c). The naive guess is offset by $0.025$, $0.049$ and $0.035$\,dex, with a slightly larger scatter of $0.030$, $0.071$ and $0.027$\,dex.}
    \label{fig:halos}
\end{figure*}

In Fig.~\ref{fig:halos}, we show the halo-by-halo prediction of the \textit{halo mass}, $M_{200m}$, enclosed in a sphere whose mean density is $200$ times the mean matter density, the \textit{halo concentration}, estimated as $c \equiv r_{90}/r_{25}$, the ratio of the radii enclosing $90\%$ and $25\%$ of the halo mass, and the \textit{maximum circular velocity}, $v_{\max} \equiv \max_r \sqrt{GM(<r)/r}$. 

Mass is the primary variable that determines how galaxies populate dark matter halos, whereas concentration and $v_{\max}$ probe the internal structure of a halo, which retains a memory of its assembly history. These are therefore most likely to show a mismatch between the growth histories of the different dark energy models and the predictor. We use $r_{90}/r_{25}$ instead of a Navarro--Frenk--White (NFW)~\citep{NFW} fit because the fits were unstable for the smallest halos we consider.

The matched models show negligible offsets: the median mass, concentration and $v_{\max}$ of their halos differ from the predictor's by at most $0.003$\,dex. The naive guess, however, is biased towards higher masses, concentrations and $v_{\max}$, by $0.025$, $0.049$ and $0.035$\,dex respectively.

Individual halos are nonetheless different. The scatter over matched pairs (its standard deviation) is highest for concentration: at most $0.022$, $0.065$ and $0.023$\,dex in mass, concentration and $v_{\max}$ for the matched models, and only slightly larger, $0.030$, $0.071$ and $0.027$\,dex, about its offset, for the naive guess. Interestingly, we find that the scatter is very similar for all matched models even though their expansion histories deviate from the predictor differently. Note that we do not separate this scatter from the contribution of the butterfly effect: the chaotic nature of gravitational dynamics amplifies any small difference between two simulations into differences between individual halos \citep{Genel_2019}. This may especially affect our concentration measurements of the low-mass halos that contain only $\sim 100$ particles. 

\subsection{Galaxy Clustering}

What surveys observe as tracers of the large-scale structure is light emitted by galaxies, which form inside collapsed halos. How galaxies populate those halos depends primarily on their mass and, to a lesser degree, on secondary properties related to the halo's assembly history such as concentration and environment. Moreover, distances to galaxies are inferred from redshifts, so the observed clustering is distorted along the line of sight by the galaxies' peculiar velocities. Galaxy clustering therefore combines all the ingredients discussed so far: the abundance, positions and internal structure of halos and their velocities.

To populate our dark-matter-only simulations we use the halo occupation distribution (HOD) model of Zheng et al.~\citep{Zheng_2007} with environmental assembly bias as implemented in halotools~\citep{halotools}. The environmental bias makes the test more stringent, since the predictor has to reproduce not only halos and their masses but also their surroundings. A halo of mass $M \equiv M_{200m}$ hosts either zero or one central galaxy, drawn from a Bernoulli distribution with mean

\begin{equation}
    \langle N_\text{cen} (M) \rangle = \frac{1}{2} \left[1 + \mathrm{erf}\left( \frac{\log M - \log M_\text{min}}{\sigma_{\log M}} \right) \right],
\end{equation}
where $M_\text{min}$ is the mass at
which half of the halos host a central galaxy, and $\sigma_{\log M}$ sets the width of the transition from zero to one central. Each central galaxy is placed at the center of its host halo and inherits the core velocity of the halo. The number of satellites is sampled from a Poisson distribution with mean

\begin{equation}
    \langle N_\text{sat} (M) \rangle = \left( \frac{M - \kappa M_\text{min}}{M_1} \right)^\alpha \langle N_\text{cen} (M) \rangle
\end{equation}
for $M > \kappa M_\text{min}$, and zero otherwise. Here $\kappa M_\text{min}$ is the cutoff mass below which halos host no satellites, $M_1$ is the
normalization, such that a halo of mass $M_1 + \kappa M_\text{min}$ hosts one satellite on average, and $\alpha$ is the slope of the satellite occupation
at high mass. The factor $\langle N_\text{cen} \rangle$ suppresses satellites in halos that are unlikely to host a central. Satellite positions are sampled from the halos' NFW profiles with a hard cut at $R_{200m}$, and their velocities are constructed from the host's core velocity with an additional Gaussian draw with zero mean and isotropic Jeans dispersion. 

The environment of each halo is the summed mass of all other hosts down to $N = 20$ particles whose centers are within $5 \, h^{-1} \,\text{Mpc}$. The halos are ranked by it: halos above the median have their mean central and satellite occupations raised by $A_\text{cen} \min\left[ \langle N_\text{cen} \rangle, 1 - \langle N_\text{cen} \rangle \right]$ and $A_\text{sat} \langle N_\text{sat} \rangle$, respectively. The halos that are below the median have the means lowered by the same amount.
We target the number density $\bar n = 2.8\times10^{-4}\,h^{3}\,{\rm Mpc}^{-3}$ and satellite fraction $f_\text{sat} = 0.10$. We fix $\sigma_{\log M} = 0.283$, $\kappa = 0.93$, $\alpha = 1.09$, $A_\text{cen} = 0.014$ and $A_\text{sat} = 0.115$, and fit $\log M_\text{min} = 13.35$ and $\log M_1 = 14.55$ (masses in $h^{-1}\,M_\odot$) once, on the predictor. All simulations are populated with these parameters and never refitted, so that halo differences propagate into the galaxies instead of being absorbed by the fit.

\begin{figure*}
    \centering
    \includegraphics[width=\textwidth]{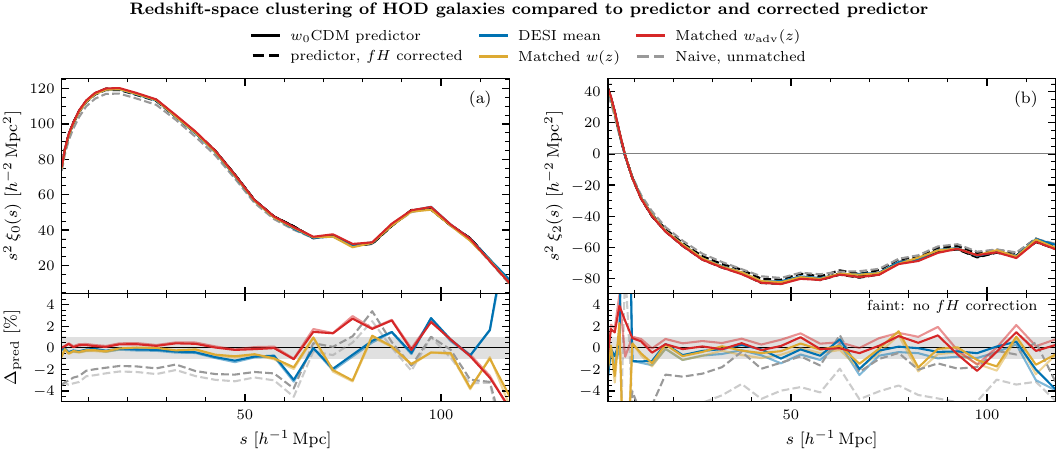}
    \caption{\textbf{Redshift-space clustering of HOD galaxies matches even at small scales.} Each box is populated with HOD galaxies 100 times, and the clustering measurements are averaged over these realizations. Faint curves show the redshift-space clustering (monopole and quadrupole) when no velocity correction is applied; solid curves show the same results when the $fH$ correction is used. Clustering errors stay within a few percent across all scales, except for the largest scales in the monopole and the smallest scales in the quadrupole, where both statistics approach zero, so relative errors get amplified.}
    \label{fig:rsd_galaxies}
\end{figure*}

To estimate the effects of HOD sample variance, each simulation is populated $100$ times. The averaged two-point clustering measurements are presented in Fig.~\ref{fig:rsd_galaxies}, where, as in Fig.~\ref{fig:vel}, faint lines show the comparison without the velocity correction. The velocity correction effect is not so apparent here: at large scales the main source of mismatch comes from halo clustering. The errors in the monopole and quadrupole stay within a few percent, except for the bins whose values are close to zero or very noisy.

\section{Data analysis prescriptions}
\label{sec:inference}

The results above show that, at a given redshift, a $w_0$CDM cosmology selected with Eqs.~\eqref{eq:dLSS_equality} and \eqref{eq:linear_equality} reproduces the clustering of a wide range of $w(z)$ models. In this section, we demonstrate how this matching method can be combined with $w_0$CDM emulators for any summary statistic, or even at the field level, to perform parameter inference in arbitrary $w(z)$ dark energy models. We validate the approach using the syren-new emulator of the matter power spectrum, and reproduce the full $w_0w_a$ posterior from $w_0$-only simulations.

We consider a data vector measured at redshifts $z_i$ and a dark energy model $w(z;\boldsymbol{\theta}_{\rm DE})$, which can be CPL, a spline, or any other
parametrization. For each set of parameters proposed by the sampler, and for each redshift $z_i$, we:
\begin{enumerate}
    \item Compute the expansion history of the target model and its comoving distance from $z_i$ to last scattering.
    \item Numerically solve Eq.~\eqref{eq:dLSS_equality} for the constant equation of state, $w_0^{\rm pred}(z_i)$, that gives the same distance. 
    \item Numerically solve the linear growth equation in the target and in the predictor cosmology, and set the amplitude of the predictor, $A_s^{\rm pred}(z_i)$, so that Eq.~\eqref{eq:linear_equality} holds at $z_i$.
    \item Evaluate the $w_0$CDM emulator at $z_i$ with $(w_0^{\rm pred}, A_s^{\rm pred})$ and all other cosmological parameters left unchanged.
\end{enumerate}
The predictions at all redshifts are then combined at the likelihood level. Note that the predictor depends on redshift, and therefore the matching cosmology is different at each redshift. Statistics that depend on velocities would additionally need the rescaling described in Sec.~\ref{sec:velocities}.

\subsection{Matter Power Spectrum Validation}

We run a toy analysis using the syren-new nonlinear power spectrum emulator~\citep{syren-new}, a symbolic distillation of EuclidEmulator2~\citep{EuclidEmulator2}. It is based on $N$-body simulations and supports a wide range of CPL $w_0w_a$ cosmologies. Moreover, it is very fast to run and differentiable due to its symbolic nature. We use it to generate a data vector and make predictions during sampling. The analysis uses 10 power spectrum bins spanning $k \in \left[ 0.01, 1.0 \right] \, h \, \text{Mpc}^{-1}$ at each of the redshifts $z \in \left\{0, 0.5, 1.0 \right\}$.

Our surrogate analysis also samples the same $w_0w_a$ plane, but instead of calling the emulator for dynamical dark energy, we follow the above recipe based on matching to a constant dark energy $w_0^\text{pred}$.

To estimate the covariance matrix, we use the fiducial high-resolution Quijote simulations, as they have the same resolution as our boxes and were run with the same code. We make a block-diagonal covariance matrix with per-block Hartlap corrections~\citep{hartlap}, assuming that redshifts are uncorrelated, as would happen in a realistic analysis of a multi-redshift probe. We also add a $2\%$ error to the diagonal terms to account for errors in syren-new and in EuclidEmulator2, on which it was trained.

To isolate the dark energy part of the setup, we fix all other parameters: $\Omega_m = 0.3175$, $\Omega_b = 0.049$, $h = 0.6711$, $n_s = 0.9624$, $M_\nu = 0$. We pick $\sigma_8 = 0.834$, $w_0 = -0.818$, $w_a = -0.6$ as our fiducial point. The priors on dark energy parameters match the training range of the emulator. For the clustering amplitude, the training range is fixed in $A_s$, rather than in $\sigma_8$. To account for this, we set a broad prior of $\sigma_8 \in [0.5, 1.0]$ and at inference time we map it to $A_s$ with a surrogate provided by syren-new. Then, if the sampled point falls outside the $A_s$ range, the prior is set to zero on the fly. We sample the posterior with the NUTS algorithm~\citep{nuts} implemented in NumPyro~\citep{numpyro}. Four chains are initialized at the ground-truth values, followed by 1000 warmup steps for parameter adaptation in NumPyro and 3000 post-warmup sampling steps. 

\begin{figure}[H]
    \centering
    \includegraphics[width=1.0\linewidth]{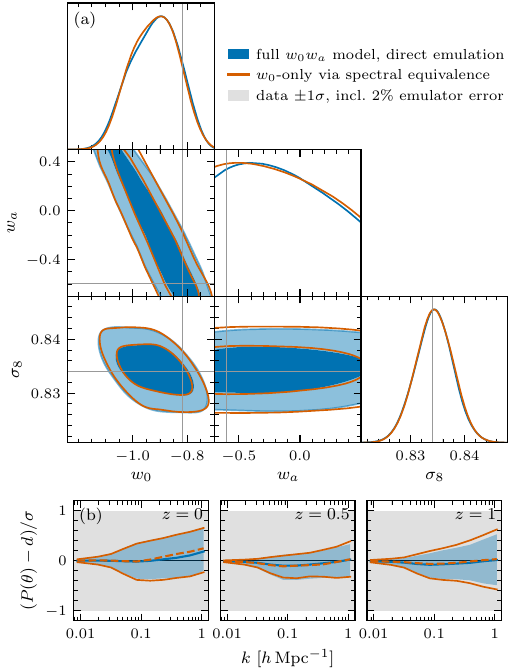}
    \caption{\textbf{Power spectrum analyses with the full $w_0w_a$ model and with a $w_0$-only model via spectral equivalence yield very similar posteriors.} (a) Comparison of the spectral equivalence $w_0$-only posterior of the power spectra at redshifts $z \in \left\{0, 0.5, 1.0\right\}$ to the ground-truth $w_0w_a$ CPL posterior. (b) Posterior predictive checks of both chains. We include a conservative 2\% error in the diagonal of the covariance matrix to account for the error of the emulator.}
    \label{fig:toy_analysis}
\end{figure}

The results of both MCMC runs are presented in Fig.~\ref{fig:toy_analysis} together with posterior predictive checks. We observe that the spectral equivalence posterior matches the original $w_0w_a$ posterior very accurately.  We also checked that all samples in the posterior were mapped to $w_0$ models that lie within the emulator range to ensure the correctness of the spectral equivalence run. To ensure the validity of the posteriors and compare them even further, we draw 300 samples from each posterior and predict the data vector with the full $w_0w_a$ model. Across all redshifts and scales, the median predictions lie within $0.25\sigma$ of the data, and their 68\% ranges extend to $\pm0.7\sigma$, with the $w_0$-only posterior closely matching the $w_0w_a$ result.

\section{Discussion}

We have shown that the clustering of matter, halos and galaxies in a universe with an arbitrary dark energy history, $w(z)$, can be predicted from a simulation with a constant equation of state, at the field level and to percent accuracy for $k \leq 1\,h\,{\rm Mpc}^{-1}$. Therefore, we can now infer the properties of any smooth dynamical dark energy model from existing simulations. In particular, large suites of simulations that span $w_0$CDM cosmologies, such as CosmoGridV1~\citep{cosmogrid} and Gower Street~\citep{gower_street}, can now be leveraged to constrain more general classes of dark energy models.

Nonetheless, our study is of an empirical nature. We understand why a mapping of this kind is possible: once time is measured by the linear growth factor, dark energy enters the dynamics of the $N$-body equations only through $\mu(\eta)$, which varies little between models. But there is still freedom in the kernel chosen for matching the two cosmologies, and there are no first principles guarantees that bound the matching error. What we do see is that a single kernel can produce a good matching to every summary statistic that we tested, and that our attempt to construct an adversarial dark energy model did not lead to any significant loss in performance. Theoretical explanations could help us close this gap, or at least inform us how the method should be applied reliably. In the meantime, any posterior can be tested by running only a few follow-up simulations to do posterior predictive checks with a full simulator, which is still much cheaper than making new simulation suites for each dark energy model of interest.

Generally speaking, spectral equivalence does not have to be limited to dark energy. We speculate that any effect that can only affect the universe through its background could be modeled by finding a $w_0\text{CDM}$ cosmology that has a matching recent expansion history. Such an expectation is very natural: one could merge curvature with dark energy into a single fluid with its own $w(z)$ and then run the same machinery to make predictions. A possible edge case this approach may face is the situation in which Eq.~\eqref{eq:dLSS_equality} cannot be solved for any $w_0$ when curvature is included, which could happen for $\Omega_k < 0$.

Our results suggest that if galaxy clustering responds to the background only through the linear amplitude and a single weighted average of the expansion history, then these are the natural parameters of the problem, and not parametrizations such as $w_0$ and $w_a$. An analysis could constrain these compressed model-independent parameters, for example as the amplitude $\sigma_8(z_i)$ and the effective equation of state $w_0^{\rm pred}(z_i)$ in each redshift bin. Any model-specific constraints can then be remapped a posteriori. The same reasoning can be applied to designing simulation suites for training emulators to ensure that they generalize to arbitrary backgrounds by construction.

Finally, while completing this manuscript, we became aware of the related work of Meng et al.\ (in prep.), who also study how a constant-$w$ cosmology can reproduce dynamical dark energy models at the field level. They write the $N$-body equations in superconformal time, $\mathrm{d}\tau \propto \mathrm{d}t/a^2$, where the background enters only through $\Omega_m a(\tau)$. This is our statement in a different time variable, since
\begin{equation}
    \mu(\eta) = \frac{\Omega_{m,0}\, a(\tau)}{\left(\mathrm{d}\ln D/\mathrm{d}\tau\right)^2},
    \label{eq:mu_tau}
\end{equation}
so that two cosmologies with the same $a(\tau)$ have the same $\mu(\eta)$ and evolve identically.

For $z>0$, Meng et al.\ keep this single $w$ and recover earlier epochs from the same simulation by matching the linear growth factor, which gives a matching cosmology at all redshifts at the cost of a small loss in accuracy. We instead re-match at every redshift, which comes at no extra cost when the predictor is an emulator (Sec.~\ref{sec:inference}). Their validation is also complementary to ours: Meng et al.\ test a grid of CPL cosmologies spanning $\Omega_m$, $w_0$ and $w_a$ at $0 \le z \le 3$ and show that the error stays small and tracks the mismatch in $a(\tau)$, whereas we test at $z=0$ but consider more general $w(z)$, including an adversarial equation of state built to break the matching method, and follow the mapping through to halos, voids and HOD galaxies in redshift space.

\section{Conclusion}

In this paper we show how to use existing constant dark energy emulators and simulation suites to construct predictions for the field-level clustering of matter, halos and galaxies, for dynamical dark energies well beyond CPL models by matching their recent average expansion histories, while keeping linear scales equal at a given redshift.
This can be thought of as splitting all dynamical dark energies into approximate equivalence classes parametrized via $w_0$. It is important to keep in mind that this equivalence is only approximate and that small-scale discrepancies exist within each class. We find that the matching error is below $0.2\%$ in the matter power spectrum and $0.26\%$ in the bispectrum for $k \leq 0.5\,h\,{\rm Mpc}^{-1}$, and reaches $0.6\%$ and $1.75\%$ at $k = 1\,h\,{\rm Mpc}^{-1}$. At the field level, we find the decorrelation between the predictor and each matched cosmology to be below $0.006\%$ up to $k = 1\,h\,{\rm Mpc}^{-1}$. These errors are subdominant to the theoretical uncertainties in modeling baryonic feedback and the galaxy--halo connection, which already limit how much information can be extracted from these scales.

We demonstrate that the spectral equivalence method is very similar to evolution mapping in spirit and can be justified using the same reasoning. However, it offers significant benefits at a mild cost: a $w_0\text{CDM}$ emulator or simulator is needed to act as an estimator of various summaries of simulations, including fields. It also does not require any correction calibration and produces a single predictor cosmology at a single redshift for every quantity of interest, including dark matter fields.

We also demonstrate the use of the spectral equivalence matching for inference in a controlled test. We compare posteriors obtained in the CPL parametrization $w_0w_a$ to those obtained through $w_0$-only emulation and spectral equivalence matching, showing that a $w_0$CDM emulator is sufficient to constrain dynamical dark energy.

\hide{\peter{Suggested conclusion structure: 1) In this paper we show how to model growth for any w(z) cosmology at the field level.  2) summary of how it works 3) Then summarise results 4) Doesn't go to crazy small scales, but who cares baryons + nonlinear galaxy bias. 5) How this fits into broader context. We need structure growth to tighten constraints and distinguish between w(z) models and other DE models. This is first step gives us full w(z)CDM model space. 6) Next step is to apply to data. Euclid, Roman, LSST, DESI full-shape}}

All of this shows that we can analyze a rich family of dynamical dark energy models without having to run massive new simulation suites or retrain existing emulators. We argue that even without a justifying theory in place, if errors are quantified reasonably well, an analysis performed with this method can always be tested with posterior predictive checks of a small number of samples using the full simulator. 

The natural next step is to apply the method to data. New datasets coming from large collaborations like Euclid, Roman, Rubin-LSST or DESI are becoming sensitive to dynamical dark energies, and one could place the first constraints on non-parametric $w(z)$ models using non-Gaussian statistics at a fraction of the cost that was expected before.
\section{Acknowledgments}
PLT acknowledges the support of the U.S. Department of Energy (DOE) award number DE-SC0011726 and NASA ROSES 21-ATP21-0050.  Portions of the data processing pipeline and scripts to make figures were drafted using Claude Code (Opus 5, Opus 5.5, Fable 5.2; Anthropic). The authors designed the analysis, reviewed all generated code, and validated its output. We would like to thank Francisco Villaescusa-Navarro, Andrei Gruzinov, Raul Angulo, Jeremy Tinker, Roman Scoccimarro and David Hogg for insightful discussions. 
\newline
{\it Authorship Contributions:} PD led the analysis and wrote the paper. CCL and PLT provided theoretical expertise and guidance.

\bibliographystyle{unsrtnat}
\bibliography{references} 

\appendix
\section{Dark energy perturbations in the Initial Conditions}
\label{ap:de_perturb}

One of the features of the \textsc{monofonIC} IC generator is that it can generate an initial configuration of particles for CPL dark energy models. We do not use this functionality in our comparisons, since this would require modifying \textsc{CLASS}, on which this code relies, to support the arbitrary equations of state $w(z)$ that we compare in the main text. 

\begin{figure*}
    \centering
    \includegraphics[width=\textwidth]{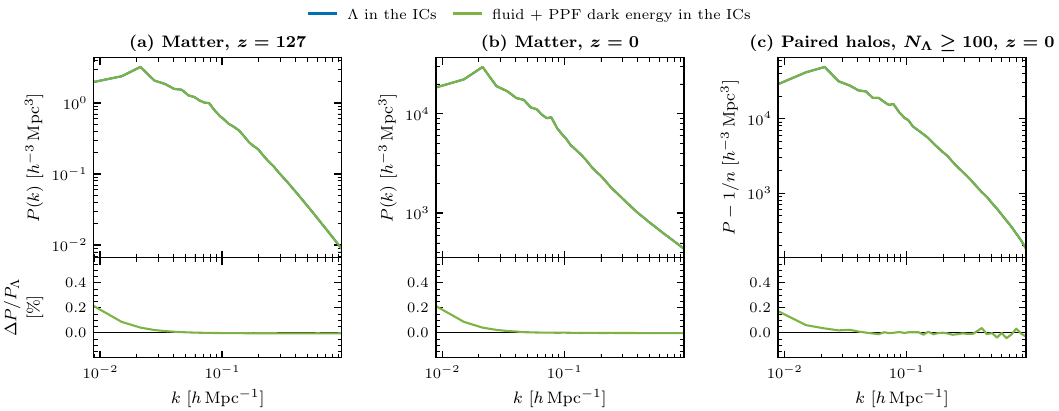}
    \caption{\textbf{Fluid + PPF treatment of dark energy in IC generation has some effects on dark matter even at late times.} (a) Clustering of dark matter at $z = 127$ for the DESI mean simulation with fluid + PPF for IC generation and with vanilla $\Lambda\text{CDM}$ ICs. Matter overclusters at the largest scales by $0.2\%$. (b) The effect of overclustered dark matter at initialization persists all the way to $z = 0$. (c) Halos with $N \geq 100$ particles from the $\Lambda\text{CDM}$ ICs are compared to their counterparts from the simulation with clustering dark energy in the ICs. This removes noise due to halos crossing the number threshold in two simulations differently.}
    \label{fig:ppf}
\end{figure*}

Generally speaking, this should not be necessary: we check that the energy density ratio $\rho_\text{DE} / \rho_m$ is below $10^{-5}$ at $z = 127$ for all matched simulations. The naive guess simulation with $w_0 = -0.42$ has a much higher value of this ratio, $\rho_\text{DE} / \rho_m \approx 4.8 \times 10^{-3}$, but this run is used only for demonstration: this cosmology is unrealistic under any data with reasonable constraining power. Moreover, at $z = 127$ dark energy should cluster at scales that are not covered by our simulated boxes, $k \ll aH/c_s \sim 2.2 \times 10^{-3}\,h\,{\rm Mpc}^{-1}$. However, we observe a small difference in the matter power spectra of the ICs generated with a $\Lambda\text{CDM}$ model and with the DESI mean cosmology. To check the impact of this mismatch at late times, we run both ICs with the DESI mean background expansion and compare matter and halo clustering at $z = 0$. To avoid noise due to cutting poorly resolved halos differently in different simulations, we estimate clustering on paired halos, with a cut imposed only in one simulation and then transferred to the other through pairing. The results are presented in Fig.~\ref{fig:ppf}. The mismatch in clustering at the largest scales, around $k \sim 10^{-2}\,h\,{\rm Mpc}^{-1}$, stays within $0.21\%$ both for matter and halos. Hence, we conclude that the choice of initial conditions does not affect our results: the resulting differences are small and subdominant to cosmic variance.

\end{document}